\documentstyle[12pt]{article}

\newcommand{\be}{\begin{equation}}
\newcommand{\bea}{\begin{eqnarray}}
\newcommand{\ee}{\end{equation}}
\newcommand{\eea}{\end{eqnarray}}

\def\theequation{\arabic{section}.\arabic{equation}}
\begin{document}
\topmargin -1cm \oddsidemargin=0.25cm\evensidemargin=0.25cm
\setcounter{page}0
\renewcommand{\thefootnote}{\fnsymbol{footnote}}
\begin{titlepage}
\begin{flushright}
hep-th/0609082 \\
DFTT-24/2006\\
\end{flushright}
\vskip .7in
\begin{center}
{\Large \bf Constructing the Cubic Interaction Vertex of Higher
Spin Gauge Fields} \vskip .7in {\large I.L.
Buchbinder$^a$\footnote{e-mail: {\tt joseph@tspu.edu.ru}}, Angelos
Fotopoulos$^{b,c}$ \footnote{e-mail: {\tt
afotopou@physics.uoc.gr}},\vskip .1in
  Anastasios C. Petkou$^b$}\footnote{e-mail: {\tt
petkou@physics.uoc.gr}}, and {\large Mirian Tsulaia$^b$}
\footnote{e-mail: {\tt tsulaia@physics.uoc.gr}} \vskip .2in {$^a$
\it Department of Theoretical Physics, Tomsk State Pedagogical
University,  634041 Tomsk,
Russia} \\
\vskip .2in {$^b$ \it Department of
Physics and Institute of Plasma Physics, University of Crete, 71003 Heraklion, Crete, Greece} \\
\vskip .2in {$^c$ \it Dipartimento di Fisica Teorica
dell'Universit\`a di Torino and Istituto Nazionale di Fisica
Nucleare, Sezione di Torino
via P.Giuria 1, I-10125 Torino, Italy} \\

\begin{abstract}

 We propose a method of construction of a cubic interaction in
 massless Higher Spin gauge theory both in flat and in AdS space-times of arbitrary dimensions.
 We consider a triplet formulation of the
 Higher Spin gauge theory and generalize the Higher Spin symmetry
 algebra of the free model to
 the corresponding algebra for the case of cubic interaction.
 The generators
 of this new algebra carry indexes which label the three
 Higher Spin fields
 involved into the cubic interaction.
 The method is based  on
 the use of
 oscillator formalism and on the Becchi-Rouet-Stora-Tyutin (BRST) technique.
 We derive  general conditions on the form of cubic interaction
 vertex and discuss the ambiguities
 of the vertex which result from  field redefinitions.
  This method  can
 in principle be  applied for constructing
 the Higher Spin interaction vertex at any order.
 Our results are a first step towards the construction of
 a Lagrangian for interacting Higher Spin gauge fields that can
 be holographically studied.

\end{abstract}

\end{center}

\vfill

\end{titlepage}

\tableofcontents

\section{Introduction}
 Classical Higher Spin
gauge theories should describe consistent dynamics of free and
interacting massive and massless particles with arbitrary values
of spin (e.g. \cite{Vasiliev:2004qz} for recent reviews of various
aspects of Higher Spin field theory). One of the leading
directions in this area is devoted to constructing a Lagrangian
formulation for Higher Spin fields in flat and AdS space-times of
arbitrary dimensions. Other than being a fascinating topic by
itself, Higher Spin field theory has attracted a significant
amount of attention due to its close relations with string - and M
- theories. Also we point out the interesting links of Higher Spin
field theory with holography ideas.

The study of higher spin (HS) gauge theories is notoriously
difficult and demanding. Already for free HS gauge fields it is
highly nontrivial to construct Lagrangians that yield HS field
equations with enough gauge invariance to remove nonphysical
polarizations - ghosts - from the spectrum. Moreover, the
requirement of gauge invariance  restricts severely  the possible
gravitational backgrounds where free fields with spin greater than
two can consistently propagate. Up to date, only constant
curvature backgrounds - Minkowski, de Sitter (dS) and Anti-de
Sitter (AdS) spaces - are known to support the consistent
propagation of HS gauge fields.

Interacting HS gauge fields are
much harder to deal with. An important landmark
was reached with the understanding of \cite{Fradkin:1986qy} --
\cite{Vasiliev:1990en} (see also \cite{Sezgin:2001zs})
that the AdS background can accommodate
consistent self - interactions of massless HS fields.
An important property of this construction is that the coupling constants of massless HS  interactions
are proportional to positive powers of the AdS radius and
therefore this picture admits no flat space-time
limit. This picture has two crucial features;
the presence of an infinite tower of
massless HS fields and nonlocality.

Studies of HS gauge fields can be grouped into two broad classes according to the particular formulation of HS gauge theory they use.
 In the Vasiliev formulation
(``Frame - like formulation") a massless
HS field with spin  $s$ is encoded into generalized
spin-- connections $\omega_\mu^{A_1,
A_2,...,A_{s-1},B_1,B_2,..,B_{s-1}}$ and the free part of the theory
is a generalization of the MacDowell -- Mansouri formulation of
gravity \cite{MacDowell:1977jt}. An
alternative formulation (``metric -- like" formulation), due to
Fronsdal, uses conventional tensor fields $\phi_{\mu_1,\mu_2,.. \mu_3}(x)$ to construct the free
gauge invariant Lagrangian both for flat space-time
\cite{Fronsdal:1978rb} and for 4-dimensional AdS
space \cite{Fronsdal:1978vb} (see \cite{Metsaev:1994ys}
 for field equations and
  \cite{Buchbinder:2001bs} -- \cite{Sagnotti:2003qa} for  Lagrangians in an arbitrary
number of dimensions). It was recently
shown that this formulation results from a partial gauge fixing
of Maxwell - like geometric equations
\cite{Francia:2002aa}--\cite{Francia:2002pt}.

In this work we undertake the first step towards constructing
explicitly the interaction vertex for HS gauge fields in AdS in
the ``metric --like" formulation. Some features of higher spin
interaction have previously been studied
 in flat space both in covariant \cite{Berends:1984rq} and in the
light-cone \cite{Bengtsson:1983pg}--\cite{Metsaev:2005ar}
formalisms. These studies have shown that for spin higher than two
a group structure for nonabelian gauge transformations fails to
exist unless one considers the full infinite tower of massless HS
fields \footnote{Two examples of  a consistent self interactions of
three fields of spin 3 was recently found in
\cite{Bekaert:2005jf}.} \cite{Bengtsson:1987jt} --
\cite{Koh:1986vg}.

The construction of consistent Higher Spin field interactions is an old
open problem of classical field theory.
 However, there exist some new motivations for the study of this problem.
The first one is the holography of HS gauge theories. It is
believed to be the appropriate framework for the holographic
description of weakly coupled gauge theories \cite{Sundborg}. In
fact, we believe that one can holographically translate the wealth
of knowledge on weakly coupled quantum field
 theories to information about HS gauge theories \cite{KP}--\cite{tassos}.
 A second motivation is stipulated by the study of the tensionless limit of
 string theory. It is widely believed that in
 the $\alpha'\rightarrow \infty$ limit the true symmetries of string
 theory will emerge \cite{Moeller:2005ez} -- \cite{Gross:1987ar} and HS
 gauge theories should play a prominent role.  It is natural to assume that a
 consistent tensionless limit can only be taken in the presence of a
 dimensionful  parameter such as space-time curvature.
  For example, curvature provides an effective tension among string-bits
  which can compensate the absence of tension providing a stringy tensionless limit.
Similar ideas appear in a number of recent
 works  \cite{Sezgin:2002rt}-- \cite{Engquist:2005yt}.

To construct the interaction vertex in AdS we use the covariant
BRST approach \cite{Pashnev:1998ti} -- \cite{Buchbinder:2004gp} of
the ``metric like formulation" to impose gauge invariance. The
theory will be formulated in terms of the HS functional - an
analogue of the string-field functional which contains an infinite
tower of massless fields with arbitrary integer spins. At the free
field level such a system describing totally symmetric reducible
representations has been considered in \cite{Sagnotti:2003qa},
\cite{Barnich:2005bn}. We extend those studies to the interacting
level.

The techniques used in the approach under consideration are
analogous in some aspects to techniques of string field theory
\cite{Neveu:1986mv}-- \cite{Gross:1986ia}. However there are some
crucial differences. Unlike string field theory, a world sheet
description of HS fields  is not known and therefore there is no
analogue of the string overlap conditions, which restrict the
argument of the cubic interaction vertex to be quadratic in the
oscillators. In our case, the interaction vertex is a general
polynomial of the oscillator and ghost variables.

We emphasize that our approach is in a sense perturbative, the
perturbation parameter being the dimensionful coupling $g$,
whose physical meaning we explain below. That is the reason why
our results in flat spacetime and in AdS do not contradict the
known no-go theorems for interacting HS gauge fields. To construct
the fully gauge invariant action, i.e. gauge invariant to all orders
in $g$,  one probably has to add
quartic and higher order interactions. We expect that the fully gauge
invariant action would contain all the known features of interacting HS theories,
such as an infinite tower of fields of all spin and possibly non-locality. Also we point out
that the symmetry algebra in HS theory is not the Virasoro one as
in string field theory. Throughout the paper we restrict to
symmetric tensor fields. This suffices if we do not include
fermions and fields with mixed symmetry \cite{Brink:2000ag}
--\cite{Alkalaev:2006rw}. We hope to address such  issues in a
future work.

The paper is organized as follows: In section \ref{FFTHS} we
review the equations of motion, the Lagrangians and their gauge
transformations describing reducible representations of the
Poincar\'e  and of the AdS group, using the triplet method for the
description of HS fields. In section \ref{TPI} we formulate the
general approach to constructing cubic interaction vertices for
massless Higher Spin fields in flat and AdS  backgrounds. We
present the main equations of the BRST analysis of the vertex,
which are used to constrain the form of the interaction vertex. We
define the coefficients of the vertex using an appropriate
expansion in ghost and "matter" oscillators. We explain  in
addition that not all possible interactions terms contain
non-trivial information about the cubic vertex. Some expansion
coefficients lead to total derivative terms, while some others
lead to "fake" interactions which can be factored out using
appropriate field redefinitions. We deal separately with the flat
and AdS cases in sections \ref{TPIF} and \ref{TPIAdS}
respectively. In \ref{TPIF} we demonstrate how one can use our
formalism to solve the equations for gauge invariance of the
vertex in the flat case, after "fake" interactions have been taken
into account. These result can be used in the sequel  to bring the
vertex into a form directly applicable on the one hand to
holography, and on the other hand to the high energy limit of
string theory. The AdS case involves some extra complications
which we discuss in section \ref{TPIAdS}. In the Appendix
\ref{ApB} we present the detailed field redefinitions formulas
used in order to factor out "fake" interactions from the cubic
vertex in section \ref{TPI}.

\setcounter{equation}0\section{Free Higher Spin Gauge Fields in
Flat and AdS space-times}\label{FFTHS}

There are many ways used in the literature to present the theory of free HS gauge fields
\cite{Vasiliev:2004qz}. We believe that one of the most elegant and clear descriptions of HS gauge fields is the one  based
on the triplet construction which we  review below.
This construction was developed in flat space in \cite{Francia:2002pt} and in AdS in
\cite{Sagnotti:2003qa}, \cite{Barnich:2005bn}. This system is
named bosonic triplet and describes the propagation of reducible
massless HS  fields in flat and AdS backgrounds. The name
``triplet`` comes about because a gauge invariant description of
massless fields with spins $s, s-2, s-4 ,...$ requires in
addition to a tensor field $\phi$ of rank $s$, the presence of
two auxiliary tensor fields. We denote them as $C$ (of rank $s-1$)
and $D$ (of rank $s-2$). After elimination of these auxiliary
fields via the gauge transformations and/or via their own
equations of motion one is left only with the degrees of freedom describing the physical polarization of
higher spin fields with spins $s,s-2,s-4$ etc.

We restrict ourselves here to the case of totally symmetric fields
on ${\cal D}$-dimensional AdS space-time. Such a tensor of rank-$s$ is the coefficient of the following state in a Fock space
\begin{equation}
\label{Phis}
|\Phi^{(s)}\rangle \ =\
\frac{1}{(s)!}\, \varphi_{\mu_1\mu_2...\mu_{s}}(x) \,
\alpha^{\mu_1 +} \ldots \alpha^{\mu_s +}  \, |0\rangle \ .
\end{equation}
We will call the vector (2.1) a higher spin functional. To
describe the triplet we introduce
  the tangent-space valued oscillators $(\alpha^a,\alpha^{a+})$,
which satisfy \be [ \alpha^a , \alpha^{b+} ] \ = \ \eta^{ab} \ ,
\quad a,b = 0,..., {\cal D}-1\,. \ee
The
oscillators $(\alpha^{\mu +},\alpha^\mu)$ are obtained
using the AdS vielbein $e^a_\mu$ and inverse vielbein $E^\mu_a$ as
  \begin{equation}   \alpha^a
=e_\mu^a \alpha^\mu
\ , \quad \alpha^\mu=E^\mu_a \alpha^a\, ,\quad [ \alpha^\mu , \alpha^{\nu +} ] \ = \ g^{\mu\nu}\,,\end{equation}
  with $g_{\mu \nu}$ being the AdS metric.
The ordinary partial derivative is now  replaced by the operator \cite{Buchbinder:2006ge},
\begin{equation}
\label{pop}
p_\mu \ = \  -\; i \, \left(
\nabla_\mu + \omega_{\mu}^{ab} \, \alpha_{\; a}^+\,
  \alpha_{ \; b} \right) \ ,
\end{equation}
where    $\omega_\mu^{ab}$ is  the spin
connection of AdS and $\nabla_\mu$ is the AdS covariant derivative.
This operator satisfies the  commutation relations
   \begin{equation}
   \label{COMU}
   D_{\mu \nu} \equiv [p_\mu,p_\nu]
= -[\nabla_\mu,\nabla_\nu]+ \frac{1}{L^2}\; (\alpha_{\; \mu}^+ \, \alpha_{\; \nu} \, -\,
\alpha_{\; \nu}^+ \, \alpha_{\; \mu}) \ , \quad [p_\mu,\alpha^{\nu +}] =0\,.
\end{equation}
The action of $p_\mu$ on the state (\ref{Phis})
gives the  AdS covariant derivative $\nabla_\mu$ as
\begin{equation}
p_\mu|\Phi^{(s)}\rangle \ =-\frac{i}{(s)!}\alpha^{\mu_1 +} \ldots
\alpha^{\mu_s +} \nabla_\mu \, \varphi_{\mu_1\mu_2...\mu_{s}}(x)
|0\rangle\,.
\end{equation}
We also write down for later use  the left action of $p_\mu$ on states
\begin{equation}
\langle \Phi^{(s)}| p_\mu =\frac{i}{(s)!}\langle 0|\,\alpha^{\mu_1 } \ldots
\alpha^{\mu_s } \nabla_\mu \, \varphi_{\mu_1\mu_2...\mu_{s}}(x)\,.
\end{equation}
Let us note also that the first term in the right hand  side in
(\ref{COMU}) gives zero when acting on states (\ref{Phis})
since the later has no free indexes. The reason behind the use
of a covariant derivative in (\ref{pop}) will be clear
when considering the case of interacting fields as we shall see below.

Next we
introduce the following operators:

-- The d'Alembertian
operator
\begin{equation} \label{lapl} l_0 \ = \
g^{\mu \nu}  p_\mu p_\nu\,,
\end{equation} which  acts on Fock-space states as
\begin{equation} l_0|\Phi^{(s)}\rangle \
=-\frac{1}{(s)!} \alpha^{\mu_1 +} \ldots \alpha^{\mu_s +}
  \Box \, \varphi_{\mu_1\mu_2...\mu_{s}}(x) \,
   \, |0\rangle \,.
\end{equation}

--The divergence operator
\begin{equation} \label{DIV}
l =  \alpha^\mu p_\mu \,,
\end{equation}
which acts on a state in the Fock space as
\begin{equation} l|\Phi^{(s)}\rangle \
=-\frac{i}{(s-1)!}\alpha^{\mu_2 +} \ldots \alpha^{\mu_s +}
  \nabla_{\mu_1} \,
\varphi^{\mu_1}{}_{\mu_2\mu_3...\mu_{s}}(x) \,  \, |0\rangle \,.
\end{equation}

--The symmetrized exterior derivative operator,
\begin{equation} \label{EXDIV}
l^+ =  \alpha^{\mu +} p_\mu\,,
\end{equation}
which acts on states in the Fock space as
\begin{equation} l^+|\Phi^{(s)}\rangle \
=-\frac{i}{(s+1)!} \alpha^{\mu +}\alpha^{\mu_1 +} \ldots
\alpha^{\mu_s +} \nabla_{\mu} \,
\varphi_{\mu_1\mu_2\mu_3...\mu_{s}}(x) \,   \, |0\rangle \,.
\end{equation}
The latter is hermitian conjugate to the operator $l$ with respect to
the scalar product
\begin{equation}
\int d^{\cal D} x \sqrt{-g} \langle \Phi^{(s)}_1||\Phi^{(s)}_2 \rangle .
\end{equation}
  It is straightforward to obtain the following commutation relations
  among the operators just introduced
\begin{equation} \label{l0}
  [ l , l^+] \ = \ \tilde{l}_0 \ ,
\end{equation}
where the modified d'Alembertian $\tilde{l}_0$ is defined as
\begin{equation} \tilde l_0 \ = l_0 \ - \
\frac{1}{L^2}\, \left( -{\cal D} \, + \, \frac{{\cal D}^2}{4} \, +
\, 4 \, M^\dagger \; M \, - \, N^2 \, + \, 2\, N \right) \ ,
\end{equation}
and
\begin{eqnarray} \label{al}
&&
[ M^\dagger \; , \; l] \ = \ -\, l^+ \ , \nonumber \\
&&[\tilde l_0 \; , \; l] \ = \ \frac{2}{L^2}\, l \, - \,
\frac{4}{L^2}N \, l \, + \,
       \frac{8}{L^2}\, l^+ \; M \ , \nonumber\\
&&[ N \; , \; l ] \ = \ -\; l  \,. \label{tripletalg1}
\end{eqnarray}
Relations (2.17) form a closed algebra which is the base for
Lagrangian construction of the massless higher spin theory in AdS
space-time. We will call it Higher Spin symmetry algebra in AdS
space.

The operators
\begin{equation} N \ = \ \alpha^{\mu +}  \alpha_{\mu} \ + \
\frac{\cal D}{2} , \quad M= \frac{1}{2} \alpha^\mu \alpha_\mu
\,, \label{Nop}
\end{equation} form an $SO(1,2)$ subalgebra of the total nonlinear algebra.
\begin{equation} \label{so21}
[N \; , \; M] \ = \ - \; 2\; M \ , \quad [M^\dagger \; , \; N] \ =
\ - \, 2\; M^\dagger \ , \quad [M^\dagger \; , \; M] \ = \ -\, N.
\end{equation}
Having this algebra at hand one can construct the corresponding
nilpotent BRST charge. There are two distinct options however,
leading to different physical results. The first option is
  to treat all operators, except $N$,  as constraints. In other words we introduce ghost and antighost variables
for each one of the operators, except of $N$. The operator $N$ can not be treated as a
constraint since it is strictly positive and it can not annihilate
any physical state.  Then, if one builds
the nilpotent BRST charge for this nonlinear algebra one arrives
to a Lagrangian description of a single Higher Spin field in  AdS \cite{Buchbinder:2001bs}.

In order to describe a
triplet on AdS one has to follow another line \cite{Sagnotti:2003qa, Barnich:2005bn} - namely to introduce
ghost and antighost variables {\it only} for the operators $\tilde
l_0$, $l$ and $l^+$. Then one constructs
a nilpotent BRST charge in the following way. First one rewrites
the second of the commutation relations (\ref{al}) in an
equivalent way
\begin{equation}
[\tilde l_0 \; , \; l] \ = \ -\frac{1}{L^2}\, (6 + 4N) l \,  \, +
\frac{8}{L^2}\,  \; M l^+\,,
\end{equation}
i.e., pushing the operators $l$ and $l^+$ to the right. Then one uses
the standard formula for the BRST charge,
\begin{equation} \label{BRSTstandard}
Q = c^A G_A - \frac{1}{2} U_{AB}{}^C c^A c^B b_C\,, \quad A,B=1,2,3\,,
\end{equation}
where $c^A=(c_0,c,c^+)$ and $b_A=(b_0,b^+,b)$  are Grassman odd ghost and antighost
variables with  ghost number $+1$ and $-1$ respectively. The ghost and
antighost variables satisfy the anticommutation relations  $\{ c^A,
b_B \} = \delta^A_B$ while  $U^{AB}_C$ are structure constants
$[G_A, G_B] = U_{AB}{}^C G_C$. However since now we have structure
functions rather than structure constants, the naive BRST charge
(\ref{BRSTstandard}) will not be nilpotent. Therefore one computes
$Q^2$ and adds compensating terms to restore nilpotence. This
procedure leads to the BRST charge \cite{Sagnotti:2003qa}
\begin{eqnarray} \label{brst} \nonumber
Q& =&c_0\; \left(\tilde{ l}_0 \, - \, \frac{4}{L^2}  N \, + \,
\frac{6}{L^2} \right)
   \, + \, c\; l^+\  \, + \, c^+ \; l
     \, - \, c^+ \; c \; b_0 \\ \nonumber
    &-& \, \frac{6}{L^2} \; c_0\; c^+ \; b
       \, - \, \frac{6}{L^2} \; c_0 \; b^+ \; c
         \, + \, \frac{4}{L^2} \; c_0 \; c^+ \; b \; N
       \, + \, \frac{4}{L^2} \;c _0 \; b^+ \;  c \; N \\
&-& \,\frac{8}{L^2} \; c_0 \; c^+ \; b^+ \; M \, + \,
\frac{8}{L^2} \; c_0 \; c \; b \; M^\dagger \, + \, \frac{12}{L^2}
\; c_0 \; c^+ \; b^+ \; c \; b \ .
\end{eqnarray}
Furthermore,  we define the ghost vacuum as
\begin{equation}
c|0\rangle_{gh}\ = \ 0 \ , \qquad b|0\rangle_{gh}\ = \ 0 \ ,
\qquad b_0|0\rangle_{gh}\ =\ 0 \ .
\end{equation}
Therefore the total vacuum is given by the product
\begin{equation}
|0 \rangle = |0 \rangle_{\alpha} \otimes |0\rangle_{gh}, \quad
\alpha^a|0 \rangle_{\alpha} =0\,.
\end{equation}
 The triplet of spin-$s$, (which involves
symmetric tensors of ranks $s$, $s-1$, and $s-2$), is  now
expressed through the following states in this enlarged Fock space
\footnote{To avoid overloading the notation, the state
$|\Phi\rangle$ will denote henceforth the triplet of spin-$s$,
unless explicitly stated otherwise.}
\begin{equation}
\label{Phifield}
|\Phi  \rangle = |\phi_1\rangle + c_0 |\phi_2\rangle
\end{equation}
where
\begin{eqnarray}
&& |\phi_1\rangle = \frac{1}{s!}\, \phi_{\mu_1 \ldots \mu_s}(x)
\alpha^{\mu_1 +} \ldots \alpha^{\mu_s +}\; |0\rangle \nonumber
\\ && \qquad + \ \frac{1}{(s-2)!}\, D_{\mu_1 \ldots \mu_{s-2}}(x)
\alpha^{\mu_1 +}
\ldots \alpha^{ \mu_{s-2} +} \, c^+ \, b^+ \; |0\rangle \ , \nonumber \\
&& |\phi_2 \rangle \ = \ \frac{-i}{(s-1)!}\, C_{\mu_1 \ldots
\mu_{s-1}}(x) \alpha^{\mu_1 +}  \ldots \alpha^{ \mu_{s-1} +} \,
b^+ \; |0\rangle \ .
\end{eqnarray}
We will call the vector (\ref{Phifield}) a higher spin functional
as well as the (\ref{Phis}). The vacuum $|0\rangle$ and the state
$|\Phi\rangle $ have ghost number zero. The corresponding gauge
transformation parameter has ghost number $-1$
\begin{equation}
|\Lambda \rangle = \frac{i}{(s-1)!} \Lambda_{\mu_1 \mu_2...
\mu_{s-1}}(x) \alpha^{\mu_1 +} \alpha^{\mu_2 +}...
\alpha^{\mu_{s-1}+} b^+ |0 \rangle\,.
\end{equation}
Then the Lagrangian, that has ghost number zero, is
\begin{equation} \label {LBRST}
{L} \ = \ \int d c_0\,\langle \Phi |\, Q \, |\Phi \rangle \ ,
\end{equation}
and it is invariant under
\begin{equation} \label{BRSTGT}
\delta | \Phi \rangle  =  Q  | \Lambda \rangle  .
\end{equation}
Now it is straightforward to
obtain the space-time Lagrangian
\begin{eqnarray} \label{LtripletBADS}\nonumber
{\cal L} &=&  \, \frac{1}{2}\ \phi \Box \phi \ + \ s\, \nabla
\cdot \phi \, C \ + \ s(s-1)\, \nabla \cdot C \, D \
  - \ \frac{s(s-1)}{2} \, D \Box D \ - \ \frac{s}{2} \,
C^2 \\ &+& \ \frac{s(s-1)}{2L^2}\, {(\phi^{'})}^2 \ - \
\frac{s(s-1)(s-2)(s-3)}{2L^2} \, {(D^{'})}^2
   \ - \ \frac{4s(s-1)}{L^2} \, D \, \phi^{'} \\
&-& \ \frac{1}{2L^2} \, \left[ (s-2)({\cal D}+s-3) \, - \, s
\right] {\phi}^2 \ + \ \frac{s(s-1)}{2L^2} \, \left[ s({\cal
D}+s-2)+6 \right]\,  D^2 \,. \nonumber
\end{eqnarray}
The equations of motion resulting from (\ref{LtripletBADS}) are
\begin{eqnarray}
&& \Box \; \phi \ = \ \nabla C + \frac{1}{L^2} \, \left\{ 8 \, g
\, D \ - \ 2 \, g \, \phi^{'} \ + \ \left[ (2-s)(3-{\cal
D}-s)-s\right] \,
\phi \right\} \nonumber \ , \\
&& C = \nabla \cdot \varphi - \nabla D \nonumber \ , \\
&& \Box \; D \ = \ \nabla \cdot C \ + \frac{1}{L^2} \left\{
[s({\cal D}+s-2) +6] D - 4 \phi^{'} - 2 g D^{'} \right\} \ .
\label{AdStriplet}
\end{eqnarray}
In the equations above $\nabla\cdot$ gives the divergence and $\nabla$ acts as the
symmetrized covariant derivative. Further, the prime ${}^{'}$ denotes the trace with respect to the AdS metric.
Taking the infinite radius limit $L\to \infty$, the Lagrangian
(\ref{LtripletBADS}) and the equations of motion
(\ref{AdStriplet}) reduce
to their flat spacetime counterparts.

The Lagrangian is invariant under the gauge transformations
\begin{eqnarray}
&& \delta \phi \ = \ \nabla \Lambda \ , \nonumber \\
&& \delta D \ = \ \nabla \cdot \Lambda \ ,  \nonumber \\
\label{adstripletgauge} && \delta C
  \ = \ \Box \ \Lambda \ + \ \frac{(s-1)(3-s-{\cal D})}{L^2}\
   \Lambda \ +
\ \frac{2}{L^2} \,  g \; \Lambda^{'} \,,  \label{deltaphidads}
\end{eqnarray}
by virtue of the standard formula for the AdS covariant derivatives acting on a
vector $\xi_\rho$
\begin{equation} \label{Cnabla}
[ \nabla_\mu, \nabla_\nu] \xi_\rho = \frac{1}{L^2}(g_{\nu \rho}
\xi_\mu - g_{\mu \rho} \xi_\nu)\,.
\end{equation}

Finally, we should make some comments regarding the spectrum.
 Massless fields of spin-$s$ in AdS  saturate the unitary bound for representations of the
AdS isometry group $O(2,{\cal D}-1)$ (see e.g. \cite{Brink:2000ag}
 for detailed discussions). Their wave equation has the
form \be \label{ubx} ( \Box \;   - \frac{1}{L^2}\left[
(2-s)(3-{\cal {\cal D}}-s) -s \right])\Phi_{\mu_1... \mu_s}(x)
=0\,,\ee where $s$ is the spin. Then, in complete analogy with the
case of flat space time one can show
  that the triplet equations
(\ref{AdStriplet})
  correctly reproduce
the unitary bound for all physical modes i.e., after the
diagonalisation of the equations and gauge transformations we
obtain the propagation of massless fields with spins $s, s-2 ...$
and proper unitary bound for each of them separately. Next, imposing
``by hand`` the extra condition
\begin{equation} \label{extra}
\phi^{'} = 2D\,,
\end{equation}
 one can completely eliminate the lower spin fields from the triplet equations and one
arrives to the so-called Fronsdal equations in AdS
\cite{Fronsdal:1978vb}. Note that after imposing (\ref{extra}) the
parameter of  gauge transformations is no more unrestricted, but
rather satisfies the condition $\lambda^{'}=0$. This extra
condition can be obtained, after partial gauge fixing
\cite{Buchbinder:2001bs}, \cite{Pashnev:1998ti} as an equation of
motion from a larger Lagrangian which contains some additional
auxiliary fields. To get a formulation for free HS theory in flat
space it is sufficient to tend the parameter L to infinity  in all
relations corresponding to AdS space.

\setcounter{equation}0\section{Method of Constructing the Cubic
Vertex for Higher Spin Gauge Fields}\label{TPI}

In this section we discuss a general construction of the cubic HS
vertex which is based on generalization of the BRST method mainly
used earlier only in free HS theory. This approach is analogous in
some aspects to vertex construction in string field theory,
however, as we have pointed out earlier, in our case there exists
no analog of the overlap conditions on the three-string
interaction vertex that would strongly restrict its form. In the
case of interacting massless HS  fields the only guiding principle
is gauge invariance which manifests itself in the requirement of
BRST invariance of the vertex.

There is one crucial point regarding interacting HS fields.
It appears that a length parameter is
necessary for the construction of the interaction vertex,
such that the latter has the right dimensions.
For HS field in flat space there is no obvious candidate
 for this length parameter. One possibility would be
to consider HS gauge fields emerging at the tensionless limit of  string
theory, in which case the role of the above mentioned parameter is played
by the inverse of the string tension $\alpha^\prime$.
On the other hand, for HS fields in curved space-times such a
dimensionful parameter is naturally
given by the inverse curvature. In particular, in the case of HS gauge fields on AdS
space-times this parameter is naturally associated with the AdS radius $L$.
Notice that the zero radius limit of such a construction is the large-curvature limit.

After these remarks  we will proceed along the  lines of
\cite{Bengtsson:1987jt}. We wish to construct the
most general cubic vertex that includes
both the case when all interacting fields have the
same spin (self interaction) as well as
the case when the interacting fields are different.
 For that we use three copies of the Higher
Spin functional defined in (\ref{Phifield}) as $|\Phi_i\rangle$,
$i=1,2,3$. If we studied the quartic vertex we would use four
copies of the Higher Spin  functional $|\Phi \rangle$ and etc.The
tensors fields in $|\Phi_i\rangle$ are all at the same space-time
point. Then, the $|\Phi_i\rangle$ interacting among each other are
expanded in terms of the set of oscillators $\alpha^{i+}_\mu,
c^{i+}$ and $b^{i+}$
\begin{equation}
[\alpha_\mu^i, \alpha_\nu^{j,+} ] = \delta^{ij} g_{\mu \nu},
\quad \{ c^{i,+}, b^j \} = \{ c^i, b^{j,+} \} = \{ c_0^i , b_0^j \} = \delta^{ij}\,,
\end{equation}
in complete analogy to the free field case.
 The BRST charge of our construction consists of three copies of the free BRST change. The full interacting Lagrangian can be
written as \cite{Neveu:1986mv} -- \cite{Gross:1986ia}
\begin{equation} \label {LIBRST}
{L} \ = \ \sum_i \int d c_0^i \langle \Phi_i |\, Q_i \, |\Phi_i
\rangle \ + g( \int dc_0^1 dc_0^2  dc_0^3 \langle \Phi_1| \langle
\Phi_2|\langle \Phi_3||V \rangle + h.c)\,, \end{equation} where
$|V\rangle$ is the cubic vertex and $g$ is a dimensionless
coupling constant\footnote{Each term in the Lagrangian
(\ref{LIBRST}) has length dimension $- {\cal D}$. This requirement
holds true for each space-time vertex contained in (\ref{LIBRST})
after multiplication by an appropriate power of the length scale
of the theory, as discussed before.}.

It is straightforward to show that the Lagrangian (\ref{LIBRST}) is
 invariant up to terms of order $
g^2$ under the nonabelian gauge transformations
\begin{equation}\label{BRSTIGT1}
\delta | \Phi_1 \rangle  =  Q_1 | \Lambda_1 \rangle  - g \int
dc_0^2 dc_0^3[(  \langle \Phi_2|\langle \Lambda_3| +\langle
\Phi_3|\langle \Lambda_2|) |V \rangle] + O(g^2)\,,
\end{equation}
\begin{equation}\label{BRSTIGT2}
\delta | \Phi_2 \rangle  =  Q_2 | \Lambda_2 \rangle  - g \int
dc_0^3 dc_0^1[(  \langle \Phi_3|\langle \Lambda_1| +\langle
\Phi_1|\langle \Lambda_3|) |V \rangle] + O(g^2)\,,
\end{equation}
\begin{equation}\label{BRSTIGT3}
\delta | \Phi_3 \rangle  =  Q_3 | \Lambda_3 \rangle  - g \int
dc_0^1 dc_0^2[(  \langle \Phi_1|\langle \Lambda_2| +\langle
\Phi_2|\langle \Lambda_1|) |V \rangle] + O(g^2)\,,
\end{equation}
provided that the vertex $V$ satisfies the BRST invariance condition
\begin{equation}\label{VBRST}
\sum_i Q_i |V \rangle=0\,.
\end{equation}
The gauge transformations (\ref{BRSTIGT1}) -- (\ref{BRSTIGT3}) are nonlinear
deformations
of previously considered abelian gauge transformations. We assume here
that the tensor fields obtained  after the expansion of the $|\Phi_i\rangle$
functionals in terms of the oscillators $\alpha^{i+}_\mu$ are
different from each
other. One can consider
cases when two or all three HS functionals contain the
same tensor fields. We
expect that in such cases the general interaction vertex will exhibit
additional symmetry properties.

In order to ensure zero ghost number for the Lagrangian, the cubic
vertex must have ghost number $3$. We make the following ansatz
for the cubic vertex
\begin{equation}\label{Vertex1}
| V \rangle= V |- \rangle_{123} \,
\end{equation}
where the vacuum $|-\rangle $, with ghost number $3$ is defined
as the product of
the individual Hilbert space ghost vaccua
\begin{equation}\label{Defghost}
|-\rangle_{123}= c^1_0 c^2_0 c^3_0\ |0 \rangle_{1} \otimes |0
\rangle_{2} \otimes |0 \rangle_{3}\,.
\end{equation}
The function $V$ has  ghost number $0$ and it is a function of the
rest of the creation operators as well as of the operators
$p_\mu^i$. In String Field Theory the rhs of (\ref{Defghost}) is
multiplied by $\delta^{\cal D}(\sum_i p_i)$ which imposes momentum
conservation on the three string vertex. In our case the analogous
constrain is to discard total derivative terms of the lagrangian
which is certainly true for flat and AdS space-times. So in what
follows we will impose "momentum" conservation in the sense
described above.

The condition of BRST invariance (\ref{VBRST})  does not
 completely fix the cubic vertex. There is an enormous
freedom due to Field Redefinitions (FR) just like in any field
theory Langrangian. It is clear in the free theory case that
any FR of the form
\begin{equation}\label{FREx}
\delta \Phi_i= F(\Phi_i)\,,
\end{equation}
gives a gauge equivalent set of equations of motion for the fields
$\Phi_i$.
Lagrangians obtained from the free one after the field redefinition
(\ref{FREx}) yield additional ``fake interactions`` and should be discarded.
For the interacting case at hand we see, from
(\ref{VBRST}), that the modified gauge variation (\ref{BRSTIGT1}) --(\ref{BRSTIGT3})
can only determine the cubic vertex up to $\tilde{Q}$-exact cohomology
terms:
\begin{equation}\label{VFR}
\delta |V \rangle= \tilde{Q} |W \rangle\,,
\end{equation}
where $\tilde{Q}=\sum_i Q_i$ and $|W \rangle$  is a state with total ghost
charge $2$. We will see in what follows that this FR
freedom can lead into major simplifications for the functional
form of the vertex.

Next, we expand  the vertex operator $|V\rangle$ and the
function $|W \rangle$ in terms of ghost variables or equivalently
in terms of the following two ghost  quantities with ghost number
zero
\begin{equation}\label{Defab}
\gamma^{ij,+}=c^{i,+} b^{j,+}, \ \ \ \ \beta^{ij,+}=c^{i,+} b^j_0\,.
\end{equation}
These are $3\times 3$ matrices with no symmetry properties.
For the cubic vertex we have the expansion
\begin{eqnarray}\label{VExp}
&|V\rangle= \Bigl\{X^1 + X^2_{ij} \gamma^{ij,+}+ X^3_{ij} \beta^{ij,+} +
X^4_{(ij);(kl)}\gamma^{ij,+}\gamma^{kl,+}+
X^5_{ij;kl}\gamma^{ij,+}\beta^{kl,+}+ \nonumber \\
&+X^6_{(ij);(kl)}\beta^{ij,+}\beta^{kl,+}
+X^7_{(ij);(kl);(mn)}\gamma^{ij,+}\gamma^{kl,+}\gamma^{mn,+}+
X^8_{(ij);(kl);mn}\gamma^{ij,+}\gamma^{kl,+}\beta^{mn,+}+ \nonumber \\
&+X^9_{ij;(kl);(mn)}\gamma^{ij,+}\beta^{kl,+}\beta^{mn,+}+
X^{10}_{(ij);(kl);(mn)}\beta^{ij,+}\beta^{kl,+}\beta^{mn,+}\Bigl\}
|-\rangle_{123}\,,
\end{eqnarray}
since  the function  $V$  in (\ref{Vertex1}) has ghost number
zero.
  In our notation we put in parentheses pairs of
indices which are symmetric under mutual exchange. For example,
$X^4_{(ij);(kl)}$ is symmetric under $(ij) \leftrightarrow  (kl)$.
The coefficient $X^4_{(ij);(kl)}$ is also antisymmetric under $ i
\to k$  since $ \{c^{i,+}, c^{k,+} \} =0$
but we have not indicated these symmetries in order
to avoid clustering notation.

In a similar manner we have the following expansion:
\begin{eqnarray}\label{WExp}
&|W\rangle_{123}= \Bigl\{W_i^1b^{i,+} + W_i^2 b^i_{0}+ W^3_{i;jk}b^{i,+}
\gamma^{jk,+}+ W^4_{i;jk}b^{i,+} \beta^{jk,+}+ W^5_{i;jk}b^i_{0}
\beta^{jk,+}+
  \nonumber \\
&W^6_{i;(jk);(lm)}b^{i,+}\gamma^{jk,+}\gamma^{lm,+}+W^7_{i;jk;lm}b^{i,+}\gamma^{jk,+}\beta^{lm,+}+
W^8_{i;(jk);(lm)}b^{i,+}\beta^{jk,+}\beta^{lm,+}+ \nonumber \\
&W^9_{i;(jk);(lm)}b^i_{0}\beta^{jk,+}\beta^{lm,+}
+W^{10}_{i;(jk);(lm);pn}b^{i,+}\gamma^{jk,+}\gamma^{lm,+}\beta^{pn,+}
+
\nonumber \\
&W^{11}_{i;jk;(lm);(pn)}b^{i,+}\gamma^{jk,+}\beta^{lm,+}\beta^{pn,+} +
W^{12}_{i;(jk);(lm);(pn)}b^{i,+}\beta^{jk,+}\beta^{lm,+}\beta^{pn,+}\Bigl\}|-\rangle_{123}\,.
\end{eqnarray}
for the FR functional $W$.

\setcounter{equation}0\section{The Cubic Vertex in Flat Space-time}\label{TPIF}

We consider HS fields in flat space-time first. Each component of
the vertex in (\ref{VExp}) has an oscillator expansion in terms of
matter oscillators $\alpha^{i,+}_{\mu}$ and derivatives $p_\mu^i$,
where the latter act to the left. As we have done throughout the
paper we will restrict our study to the case of totally symmetric
massless higher spin fields  and therefore we have only to
consider three different sets of oscillators and momenta.

\subsection{Flat Space-Time Generators and Their Algebra for the Interacting
Case} The interaction vertex glues together three Hilbert
spaces and for this reason it is convenient to define, in complete
analogy to the free case, the following generators
\begin{eqnarray}\label{VGen}
&l^{ij}=\alpha^{\mu i} p_{\mu}^j, \ \ l^{ij,+}=\alpha^{ \mu , i+}
p_{\mu}^j, \ \ l_0^{ij}=p^{\mu i} p_{\mu}^j \,,\nonumber \\
&M^{ij}= \frac{1}{2}\alpha^{\mu i}\alpha_{\mu}^{j}, \ \ M^{ij,+}=
\frac{1}{2}\alpha^{  \mu ,i +}\alpha_{\mu}^{j +} \,,\nonumber \\
&N^{ij}= \alpha^{ \mu ,i +} \alpha_{\mu}^j +
\delta^{ij}\frac{{\cal D}}{2}\,.
\end{eqnarray}
We see that generators (\ref{VGen}) are indexed by integers $i, j
=1,2,3$. The three values for i and j originate from the fact that
we consider three field interaction. In general case of n-field
interaction , we should take the same generators with $i,j = 1,2
.., n$. Using the generators above one can build all possible
interaction terms between symmetric Higher Spin fields. Therefore
our ansatz for the vertex is that of the most general polynomial
made out from the operators $l^{ij}_0$, $l^{ ij,+}$ and
$M^{ij,+}$. This corresponds to the usual derivative expansion for
the vertex, since the operators $l_{0}^{ij}$ have dimensions
[Length]$^{-2}$ and the operators  $l^{ij,+}$, $\beta^{ij,+}$ have
dimension [Length]$^{-1}$. To make sense of such an expansion one
needs to introduce a physical length parameter. In flat
space-times it is not clear where does such a length scale may
come from, nevertheless the hope is that it
 would be connected to the length scale of a fundamental theory such as string or M-theory.

The commutator algebra of the operators in (\ref{VGen}) is:
\begin{eqnarray}\label{VGenA}
&[l^{ij}, l^{kl,+}]= \delta^{ik} l_0^{jl},
\ \ [N^{ij},l^{kl}]=-\delta^{ik}l^{jl}\,, \nonumber \\
&[M^{ij,+},l^{kl}]= -\frac{1}{2} ( \delta^{jk}
l^{il,+}+\delta^{ik}l^{jl,+}), \ \
[N^{ij},M^{kl}]=-(\delta^{ik}M^{jl}+\delta^{il}M^{kj})\,, \nonumber
\\
&[M^{ij},M^{kl,+}]=- \frac{1}{4}
(\delta^{jk}N^{il}-\delta^{jl}N^{ik}-\delta^{ik}N^{jl}-\delta^{il}N^{jk})\,.
\end{eqnarray}
Algebra (\ref{VGenA}) generalizes the algebra of generators of the
free HS theory and can be called the symmetry algebra of
interacting HS theory. It is obvious that the diagonal subalgebra
of (\ref{VGenA}) \footnote{This algebra consists of generators
$(l_0^{ii}, l^{ii},l^{ii,+}, M^{ii,+}, M^{ii}, N^{ii,+} )$ for
$i=1,2,3$.} consists of three copies of the algebra presented in
(\ref{al}), (\ref{so21}).

Let us consider the constrains imposed by momentum conservation on
the vertex. Clearly, not all generators in (\ref{VGen}) are
linearly independent once we consider the operatorial equation
$\sum_i p^\mu_i=0$, which means that we omit total derivatives, as
discussed in section \ref{TPI}. A convenient set of linearly
independent generators is the following:
\begin{eqnarray}\label{LIVGen}
&l_0^{ij}=(l_0^{11},l_0^{22},l_0^{33})=(l_0^{1},l_0^{2},l_0^{3})
\nonumber \\
\nonumber \\
&l^{ij,+}= (l^{1,+},I^{1,+},l^{2,+},I^{2,+},l^{3,+},I^{3,+}), \nonumber
\\
&l^{i,+}=l^{ii,+}, \ \ I^{1,+}=\alpha^{ \mu, 1
+}(p_{\mu}^2-p_{\mu}^3) ,  \nonumber
\\
&I^{2,+}=\alpha^{ \mu, 2 +}(p_{\mu}^3-p_{\mu}^1) \ \
I^{3,+}=\alpha^{ \mu, 3 +}(p_{\mu}^1-p_{\mu}^2)
\nonumber \\
&M^{ij,+}= (M^{11,+},M^{22,+},M^{33,+},M^{12,+},M^{13,+},M^{23,+})
\end{eqnarray}

Based on  the above analysis we can write the most general form of
the expansion coefficients $X^l_{(\dots)}$:
\begin{eqnarray}\label{XEXP}
&X^{l}_{(\dots)}=
X^{l}_{n_1,n_2,n_3;m_1,k_1,m_2,k_2,m_3,k_3;p_1,p_2,p_3,r_{12},r_{13},r_{23}(\dots)}
\nonumber \\
&(l_0^1)^{n_1} \dots (l^{+,1})^{m_1} (I^{+,1})^{k_1} \dots
(M^{+,11})^{p_1}\dots (M^{+,12})^{r_{12}} \dots
\end{eqnarray}
An analogous expansion can be written for $W^l_{(\dots)}$ as well.
It is interesting to point out that the expansion in $l^{ij},
l^{ij,+}, l^{ij}_{0}$ is an expansion in powers of space-time
derivatives. Therefore it is naturally to expect that the cubic
vertex as well as any interaction vertex will contain higher
space-time derivatives. In principle, this circumstance allows us
to develop a perturbation scheme for finding the vertex keeping in
the expansion derivatives up to some fixed order.

An easy way to recognize  interactions obtained from the free
Lagrangian (``Fake interactions") due to field redefinitions is
the following \cite{Berends:1984rq}. Fake interactions vanish for
the fields obeying free equations of motion - which is another way
to state that the only nontrivial interaction are in the cohomology of
the BRST charge $\tilde{Q}$. Fake interactions can in principle be
completely eliminated using the
field redefinitions in (\ref{WExp}). In Appendix \ref{ApB} we
demonstrate how FR can bring the matrix element $X^n_{\dots}$ of
the vertex in a convenient form, both for analyzing the equations
(\ref{VEQN}) and for writing  the Lagrangian in  a simpler form.

\subsection{BRST Invariance Constraints for the Cubic Vertex}

Using the explicit form of the BRST charges:
\begin{equation}\label{QBRST}
Q^i=c_0^i l_0^i+ c^i l^{i, +}+ c^{i,+}
l^i-c^{i,+}c^ib_0^i\,,\,\,\,\,\,\mbox{(no sum)}
\end{equation}
and equations (\ref{VBRST}), (\ref{XEXP}) we arrive to the
following set of equations:
\begin{eqnarray}\label{VEQN}
&c^{i,+}[l^i X^1-l^{s,+} X^2_{is}-l_0^sX^3_{is}]=0 \nonumber \\
&c^{i,+} \gamma^{jk,+}[l^iX^2_{jk}-2l^{s,+}X^4_{(is);(jk)} -l_0^s X^5_{jk;is}]=0
\nonumber \\
&c^{i,+} \beta^{jk,+}[-\delta_{jk}X^2_{ij}+ l^i
X^3_{jk}-l^{s,+}X^5_{is;jk}-2l_0^sX^6_{(is);(jk)}]=0 \nonumber
\\
&c^{i,+}\gamma^{jk,+} \gamma^{lm,+}[l^iX^4_{(jk);(lm)}- 3l^{s,+}X^7_{(is);(jk);(lm)}
-l_0^sX^8_{(jk);(lm);is}]=0 \nonumber \\
&c^{i,+} \gamma^{jk,+} \beta^{lm,+}[-2\delta_{lm}X^4_{(il);(jk)}+
l^iX^5_{jk;lm}-2l^{+s}X^8_{(is);(jk);lm}-2l_0^sX^9_{jk;(is);(lm)}]=0
\nonumber \\
&c^{i,+} \beta^{jk,+} \beta^{lm,+}[-\delta_{jk} X^5_{ji;lm}+ l^i
X^6_{(jk);(lm)}-l^{s,+}X^9_{is;(jk);(lm)}-3l_0^sX^{10}_{(is);(jk);(lm)}]=0\,.
\end{eqnarray}
To simplify the analysis of these equations we define the
operator:
\begin{equation}\label{DefN}
\tilde N=\alpha^{\mu, i+} \alpha^i_{\mu}+b^{i,+} c^{i}+
c^{i,+}b^i\,.
\end{equation}
This operator commutes with the BRST charges $Q_i$ and its
eigenvalues count the degree of the $X^l_{(...)}$s in
the $\alpha_\mu^{i,+}$ oscillator expansion. Namely, as it can be seen from the equation
(\ref{VExp}), if the degree of the coefficient $X^1$ in oscillators
$\alpha_\mu^{i ,+}$ is $K$, then
the rest of the coefficients have the following degrees in the oscillators $\alpha_\mu^{i,+}$
$$ X^1(K), \quad
X^2(K-2), \quad X^3(K-1),  \quad X^4(K-4), \quad X^5(K-3),$$ $$
X^6(K-2), \quad  X^7(K-6), \quad  X^8(K-5), \quad X^9(K-4), \quad
X^{10}(K-3)
$$
For example
  the first equation has degree $K-1$, since $l^{ij}$
reduces the value of $K$ by one, $l^{ij,+}$ increases it by one
and $l_0^{ij}$ leaves it unchanged.

There is yet another number which can be used in a way similar to
$K$. Namely if a term in  the expansion of $V$ has powers of operators
$l_0^{ij}, l^{ij,+}, M^{ij,+}, \gamma^{ij,+}$ and $\beta^{ij,+}$ equal
to $s_1, s_2, s_3, s_4$ and $s_5$ respectively, then the total
number $s= s_1+ s_2 +s_3 +s_4 +s_5$ is unchanged under the action
of the BRST charge.

The above observations   can be used to classify
  equations (\ref{QBRST}) by their degree $K$ and by the number $s$. This means
  that the vertex can be expanded in a sum of contribution with fixed degrees $K$ and $s$ as
  \begin{equation}
  |V\rangle =\sum_{K,s}|V(K,s)\rangle\,.
  \end{equation}
Therefore  the  equation
(\ref{VBRST}) can be split into the infinite sets of equations
\begin{equation}\label{VBRST1}
\sum_i Q_i V(K,s) =0.
\end{equation}
for each value of $K$ and $s$.

\subsection{Determining the Cubic Vertex: An Example}

We will now show how the first of equations (\ref{VEQN}) can be
solved resulting in recursive relations which determine the
expansion coefficients $X^2_{(\dots)}$ and $ X^3_{(\dots)}$ in
terms of $X^1_{(\dots)}$. Using the notation in (\ref{Trans}) we
can write the first equation of (\ref{VEQN}) in the form:
\begin{eqnarray}\label{X1EQN}
&&m_i X^1(n_i:n_i+1, m_i:m_i-1) \nonumber \\
&& - \frac{k_i}{2} \epsilon_{ikl}
(X^1(n_k:n_k+1,k_i:k_i-1)-X^1(n_l:n_l+1,k_i:k_i-1)) \nonumber \\
&&+ p_i X^1(m_i:m_i+1,p_i:p_i-1)- \frac{1}{4} \sum_l
r_{il}X^1(r_{il}:r_{il}-1,m_l:m_l+1) \nonumber \\
&&+ \frac{1}{4}\sum_l r_{il}
\epsilon_{lim}X^1(r_{il}:r_{il}-1,k_l:k_l+1) \nonumber \\
&&- \sum_s X^2_{is}(m_s:m_s+1) -\sum_s X^3_{is}(n_s:n_s+1)=0\,.
\end{eqnarray}
We have omitted all indices of the $X^{n}_{\dots}$ matrices which
are not relevant above.
  Let us look at the equation for $i=1$. Using
the specific scheme described in (\ref{Vcomp}) we observe that first
term in (\ref{X1EQN}) vanishes since it would require non-zero
matrix elements of the form $X^1_{m_i \geq 1 \dots}$ . Collecting
all term with same exponents with respect to the generators, we
find a set of equations between the various matrix elements
$X^l_{n_1,n_2 \dots;(\dots) }$. Namely terms which have
$n_i=m_i=0$  give:
\begin{equation}\label{SX11}
r_{12}X^1_{\dots k_2, k_3+1; \dots r_{12}, r_{13}-1, r_{23}}=
r_{13}X^1_{\dots k_2+1, k_3; \dots r_{12}-1, r_{13}, r_{23}}\,,
\end{equation}
with the obvious restrictions that all indices of the matrix
elements must be positive integer numbers.

We can similarly analyze all other relevant terms. In the table
bellow we summarize our results for $i=1$. The left column shows
the exponents of the generators which we factor out from
(\ref{X1EQN}) and the right one the corresponding equation between
matrix elements resulting from this process. Notice that we give
the values of the $n_i,m_i$ powers of the generators
$(l^i_0),(l^{+i})$ respectively only. All other exponents in
(\ref{X1EQN}) that are not shown here are taken  to be arbitrary
positive integer numbers. We have not included some equations
which are directly deduced by symmetry from those bellow.
\begin{equation}\label{table}
\begin{array}{|c|c|}
   \hline
   Powers \ of  \ Generators&Equations \ for \ i=1\\
   \hline
   n_i=0, m_i=0&X^1_{k_2, k_3+1;r_{12}, r_{13}-1, r_{23}}=
\frac{r_{13}}{r_{12}}X^1_{k_2+1, k_3; r_{12}-1,r_{13}, r_{23}} \\
\hline
  n_2=1, n_1=n_3=m_i=0&-k_1 X^1_{n_i=0, k_1}-X^3_{(12); n_i=0, k_1-1}=0 \\
  \hline
  n_3=1, n_1=n_2=m_i=0&-k_1X^1_{n_i=0, k_1}+X^3_{(13); n_i=0, k_1-1}=0 \\
  \hline
  m_1=1, m_2=m_3=n_i=0&p_1X^1_{m_i=0, p_1}- X^2_{(11); m_i=0, p_1-1}=0\\
  \hline
  m_2=1, m_1=m_3=n_i=0&r_{12}X^1_{m_i=0, r_{12}}+4 X^2_{(12); m_i=0, r_{12}-1}=0\\
  \hline
  n_1=1, n_{2,3}=0, m_{i}\geq 0&X^3_{(11);n_i=0,m_i\geq 0}=0 \\
\hline
   n_2=1, m_1\geq 1, n_{1,3}=m_{2,3}=0&X^3_{(12);n_i=0,m_1 \geq 1,m_{2,3}=0}=0 \\
   \hline
   any  \ two \ m_i \geq 1 &X^3_{(ij);n_i=0,any \ m_i \geq 1}=0 \\
   \hline
\end{array}
\end{equation}
Performing the same analysis for $i=2,3$ in (\ref{X1EQN}) one can
see that all $X^3_{(ij); m_i \geq 1}$ elements can be set equal to
zero. In addition the diagonal  $X^3_{ii}$ elements vanish
altogether. All other ones $X^3_{(ij); m_i=0}$ and $X^2_{ij}$ are
determined in terms of one single unknown matrix $X^1$. There is
only one constrain on $X^1$ at this level from (\ref{SX11}). The
remaining equations in (\ref{VEQN}) are very similar to the one we
just analyzed. It might be possible to determine all $X^{n\geq 2}$
matrix elements in terms of a single infinite dimensional matrix
$X^1$. However we postpone this rather lengthy analysis for a
future publication.

One can simply check that the solution given
in \cite{Koh:1986vg} emerges as a special case from the scheme described  above.
In \cite{Koh:1986vg} the function $V$ was taken to be of  the form
\begin{equation}\label{KOansatz}
V= exp(Y_{ij} l^{ij,+} + Z_{ij} \beta^{ij,+})\,.
\end{equation}
Expanding the exponential in terms of $\beta^{rs,+}$ one can see that the only nonzero
functions are  $X^1$, $X^3$, $X^6$ and $X^{10}$.
The requirement of the BRST invariance of the vertex
gives us the following conditions on the coefficients  $Y^{rs}$ and $Z^{rs}$
\begin{equation}\label{KOsolution}
Z_{i,i+1}+Z_{i,i+2}=0
\end{equation}
\begin{equation}
Y_{i,i+1}= Y_{ii}-Z_{ii} -1/2(Z_{i,i+1}-Z_{i,i+2})
\end{equation}
\begin{equation}
Y_{i,i+2}= Y_{ii}-Z_{ii} +1/2(Z_{i,i+1}-Z_{i,i+2})
\end{equation}

Nevertheless, it is instructive to divide the exponent into two
parts and use the basis (\ref{LIVGen}):
\begin{equation}\label{D1}
\Delta_1= \tilde{Y}_{r} I^{r,+} + \hat{Z}_{rs}\beta^{rs,+}
\end{equation}
\begin{equation}\label{D2}
\Delta_2= \tilde{Y}_{rr} l^{r,+} + Z_{rr}\beta^{rr,+}
\end{equation}
where compared to the basis in (\ref{KOansatz}), $\hat{Z}_{rs}$ is
the off-diagonal part of $Z_{rs}$,
\begin{eqnarray}\label{KOdefs}
&&\tilde{Y}_r=\frac{1}{2}\epsilon_{rkl} Y^{kl} \nonumber \\
&&\tilde{Y}_{rr}= Y_{rr} -\frac{1}{2}(Y_{rs}+Y_{sk}) \qquad
(r,s,k) \, \,  permutations \, \, of \, \,  (1,2,3)
\end{eqnarray}
as one can check using (\ref{LIVGen}). In $\Delta_2$ we have
grouped all the diagonal terms of the exponent. The important
point is that the two exponentials will lead into linearly
independent conditions under BRST invariance. To prove
the BRST invariance of the full vertex.
it suffices to show that $exp(\Delta_1)$
and $exp(\Delta_2)$ are separately invariant.
Since $\tilde{Q} \Delta_1$ commutes with $\Delta_1$ we can easily show that
\begin{equation}\label{eD1con}
\tilde{Q} \ exp(\Delta_1)|-\rangle_{123}= exp(\Delta_1)\
(\tilde{Q} \ \Delta_1|-\rangle_{123})\,.
\end{equation}
This implies that BRST invariance in (\ref{eD1con}) is equivalent
to
\begin{equation}\label{D1con}
\tilde{Q} \ \Delta_1|-\rangle_{123}= \sum_i c^{i,+}(-\frac{1}{2}
\epsilon_{ikl}\tilde{Y}_i(l_0^k-l_0^l)-\hat{Z}_{is}
l_0^s)|-\rangle_{123}=0\,,
\end{equation}
where we have used the commutation relations (\ref{CLIVGen}). From
the condition above we find the solution
\begin{eqnarray}\label{solD1}
&&\tilde{Y}_{i}= -\frac{1}{2} \epsilon_{ikl} \hat{Z}^{kl}\,,
\nonumber \\
&&\hat{Z}_{ik}=-\hat{Z}_{il} \qquad (i,k,l) \, \,  permutations \,
\, of \, \, (1,2,3)\,.
\end{eqnarray}
In a similar manner we can show that BRST invariance of
$exp(\Delta_2)$ is equivalent to
\begin{equation}\label{D2con}
\tilde{Q} \ \Delta_2|-\rangle_{123}= \sum_i c^{i,+}(Y_{ii}
l_0^i-Z_{ii} l_0^i)|-\rangle_{123}=0
\end{equation}
with solution
\begin{equation}\label{solD2}
Z_{rr}=\tilde{Y}_{rr}
\end{equation}
We can easily verify, using the transformations in (\ref{KOdefs}),
that the conditions (\ref{solD1}) and (\ref{solD2}) are equivalent
to those of (\ref{KOsolution}). The two conditions (\ref{D1con})
and (\ref{D2con}) are obviously linearly independent in terms of
$c^{i,+}$ and $l_0^i$ and this is the reason they can be satisfied
independently. This shows that the full vertex
\begin{equation}
|V \rangle= exp(\Delta_1 +\Delta_2)|-\rangle_{123}
\end{equation}
is BRST invariant with the coefficients satisfying (\ref{solD1})
and (\ref{solD2}). Note that
$\Delta_2$ is $\tilde{Q}$-exact, modulo terms which vanish acting
on the vacuum $|-\rangle_{123}$. This means that the dependence
of the vertex on $\Delta_2$ can be eliminated via a FR. We can
easily show that
\begin{equation}\label{FRKO}
|W \rangle= -Y_{jj} \ b^{j,+} \ exp(\Delta_1) \ \sum_{l=0}^{\infty}
\frac{\Delta_2^l}{(l+1)!} \ |-\rangle_{123}
\end{equation}
leads to the FR
\begin{equation}
\delta |V \rangle= \tilde{Q} |W \rangle= - (exp(\Delta_2)-1)\
exp(\Delta_1) \ |-\rangle_{123}
\end{equation}
and gives $V' = exp \, (\Delta_1)$. This is the scheme
(\ref{VcompB}) we have developed in Appendix \ref{ApB} to remove
"Fake Interactions". We should point out that the specific scheme
does not actually remove all diagonal ghost terms like
$\beta^{ii,+}$. Such terms appear in the $\beta$-expansion of the exponent
to quadratic order and beyond. This is because terms like
$\beta^{ik,+}\beta^{ji,+}$ can be equivalently written as
$\beta^{ii,+}\beta^{jk,+}$. It is only the $l^{i,+}$ terms which are
removed.

Having done all of the above, it is straightforward to show that
the first equation in (\ref{VEQN}) leads to the same constrains as
in (\ref{solD2}). The ansatz $V' = exp \, (\Delta_1)$ is
a particular case of the general solution in (\ref{table}) with
\begin{eqnarray}\label{KOX}
&&X^1=exp \,(\tilde{Y}_{r} I^{+ r}), \quad X^2_{ij} =0 \nonumber \\
&& X^3_{ij}\ \beta^{ij,+}= \tilde{Z}_{ij}\ \beta^{ij,+} \ X^1
\end{eqnarray}
and the matrix elements are
\begin{equation}
X^1_{n_i=m_i=0, k_1,k_2,k_3}= \frac{1}{k_1!k_2!k_3!}
\end{equation}
The remaining equations in (\ref{VEQN}) are rather straight
forward to solve in this particular case and determine $X^6$ and
$X^{10}$ in terms of $X^1$. These agree with the expansion of
$V'$ in ghost variables $\beta^{ij,+}$.

\setcounter{equation}0\section{The Cubic Vertex on AdS}\label{TPIAdS}

To construct the vertex on AdS we use the same procedure as in the flat case, in particular we solve the same equation
(\ref{VBRST}). In this case, however, care is needed when trying to extend the
the algebra (\ref{VGenA}) to a nontrivial background.

\subsection{AdS Generators and Their Algebra in the Interacting
Case}

In order to compute the algebra it is useful to recall how various
operators defined previously  act on physical states. For example
operator $l_0^{12}= p_{\mu}^1 p_{\mu}^2 $, where $p_{\mu}$ is the
operator (\ref{pop}), acts as follows
\begin{eqnarray} \nonumber
l_{0}^{12} |\Phi_1\rangle \otimes |\Phi_2\rangle &=&
\frac{i}{(s_1)!}\alpha^{\mu_1, 1 +} \ldots \alpha^{\mu_s,1 +}
\nabla^\mu \, \varphi^1_{\mu_1\mu_2...\mu_{s_{1}}}(x) |0\rangle_1
\otimes \\ \nonumber &&\frac{i}{(s_2)!}\alpha^{\nu_1,2 +} \ldots
\alpha^{\nu_s, 2 +} \nabla_\mu \,
\varphi^2_{\nu_1\nu_2...\nu_{s_{2}}}(x) |0\rangle_2,
\end{eqnarray}
The operators $p_{\mu}^i$ act only on $i$ -th Hilbert space and
therefore
  \begin{equation} \label{COMU1} [p^i_\mu,p^j_\nu]
= \delta^{ij} \, (-[\nabla^i_\mu,\nabla^i_\nu]+  \frac{1}{L^2}\;
(\alpha_{\; \mu}^{i,+} \, \alpha_{\; \nu}^i \, -\, \alpha_{\;
\nu}^{i,+} \, \alpha_{\; \mu}^i) \, ) = \delta^{ij} D_{\mu \nu}^i\
,
\end{equation}

The other operators are defined in an analogous way. For example the
operator $l^{12} = \alpha^{\mu,1 } p_{\mu}^2$ acts as
\begin{eqnarray} \nonumber
l^{12} |\Phi_1\rangle \otimes |\Phi_2\rangle &=&
\frac{1}{(s_1-1)!}\alpha^{\mu_2,1 +} \ldots \alpha^{\mu_s,1 +}  \,
\varphi^{1 \mu}{}_{\mu_2...\mu_{s_{1}}}(x) |0\rangle_1 \otimes \\
\nonumber &&-\frac{i}{(s_2)!}\alpha^{\nu_1,2 +} \ldots
\alpha^{\nu_s,2 +} \nabla_\mu \,
\varphi^2_{\nu_1\nu_2...\nu_{s_{2}}}(x) |0\rangle_2,
\end{eqnarray}
the operator $l^{12+} = \alpha^{\mu,1 + } p_{\mu}^2$ acts as
\begin{eqnarray} \nonumber
l^{12+} |\Phi_1\rangle \otimes |\Phi_2\rangle &=& \frac{1}{(s_1)!}
\alpha^{\mu, 1 +} \alpha^{\mu_1,1 +} \ldots \alpha^{\mu_s,1 +}  \,
\varphi^{1}_{\mu_1...\mu_{s_{1}}}(x) |0\rangle_1 \otimes \\ \nonumber
&&-\frac{i}{(s_2)!}\alpha^{\nu_1,2 +} \ldots \alpha^{\nu_s,2 +}
\nabla_\mu \, \varphi^2_{\nu_1\nu_2...\nu_{s_{2}}}(x) |0\rangle_2,
\end{eqnarray}
and the operator  $M^{12}= \frac{1}{2}\alpha^{\mu,1} \alpha_{\mu}^2 $ acts as
\begin{eqnarray} \nonumber
M^{12} |\Phi_1\rangle \otimes |\Phi_2\rangle &=& \frac{1}{2}
\frac{1}{(s_1-1)!}\alpha^{\mu_2,1 +} \ldots \alpha^{\mu_s,1 +}  \,
\varphi^{1 \mu}{}_{\mu_2...\mu_{s_{1}}}(x) |0\rangle_1 \otimes \\
\nonumber &&\frac{1}{(s_2-2)!}\alpha^{\nu_2,2 +} \ldots
\alpha^{\nu_s +}  \, \varphi^2_{\mu \nu_2...\nu_{s_{2}}}(x)
|0\rangle_2,
\end{eqnarray}
The definition of the diagonal operators is the same as (\ref{lapl}),
(\ref{DIV}) and (\ref{EXDIV}).

At this point we think it is instructive to present an explicit example of
a computation. Let us compute the commutator between $l^{11}$ and
$l^{12+}$ acting on $|\Phi_1\rangle \otimes |\Phi_2\rangle$, where,
for clarity, we take $|\Phi_1 \rangle$ to be a vector and $|\Phi_2 \rangle$
to be a scalar.
\begin{eqnarray} \label{example}
&&[\alpha^{\mu, 1}p_{\mu}^1, \alpha^{\nu,1 +} p_{\nu}^2]
\,\varphi^1_\rho \alpha^{\rho,1 +}|0\rangle_1 \otimes
\varphi^2|0\rangle_2 = \nonumber \\
&&=-i\Bigl(\alpha^{\mu,
1}[p_\mu^1,\alpha^{\nu,1+}]\phi_\rho^1\alpha^{\rho,1+}+
[\alpha^{\mu,1},\alpha^{\nu,1+}]p_\mu^1\phi_\rho^1\alpha^{\rho,1+}\Bigl)
(\nabla_\nu\phi^2)|0\rangle_1\otimes |0\rangle_2\nonumber \\
&&=-\alpha^{\rho,1+}(\nabla_\nu \varphi^1_\rho)(\nabla_\nu
\varphi^2)|0\rangle_1 \otimes |0\rangle_2
\end{eqnarray}
In obtaining the above result it was crucial that $p^i_\mu$
commutes with $\alpha^{ \nu, j + }$.

Proceeding this way one obtains the algebra of operators
\begin{equation}\label{hatpdef}
l_0^{ij}=p^{\mu, i} p^j_\mu \qquad l^{ij}= \alpha^{\mu, i} p^j_\mu
\qquad l^{ij,+}= \alpha^{ \mu i, +} p^j_\mu
\end{equation}
 on AdS
  for the interacting case
\begin{equation}\label{AdSGenA}
[l^{ij}, l^{mn,+}] = \delta^{im}l_{0}^{jn} -\delta^{jn}
\alpha^{\mu m, +} D^j_{\mu \nu} \alpha^{\nu i}
\end{equation}
\begin{equation}\label{AdSGenA2}
[l^{mn}, l^{kl}] =  \delta^{nl} \alpha^{\mu m} D^j_{\mu \nu}\alpha^{\nu k}
\end{equation}
\begin{equation} \label{AdSGenA4}
[l^{ij}_0, l^{mn}] = \delta^{jn} \alpha^{\nu m} D^j_{\mu \nu}
p^{\mu, i} + \delta^{in} \alpha^{\nu m}  D^i_{\mu \nu} p^{\mu, j}
\end{equation}
\begin{eqnarray}\label{AdSGenA3}
[l_0^{ij},l_0^{kl}]&=& \delta^{jk}{p}^{\mu, i} D^j_{\mu \nu}p^{\nu
,l}+ \delta^{ik}{p}^j_\mu D^i_{\mu \nu}{p}^l_\nu
-\delta^{jl}p^{\mu, k} D^j_{\mu
\nu}p^{\nu, i} \nonumber \\
&&-\delta^{il}p^{\mu, k} D^i_{\mu \nu}p^{\nu j} +\frac{(1-{\cal
D})}{L^2}\delta^{ik}\delta^{ij}l_0^{il} -\frac{(1-{\cal
D})}{L^2}\delta^{jl}\delta^{ij}l_0^{ki}
\end{eqnarray}
supplemented by the part of the algebra (\ref{VGenA}) which
involves commutators of $M^{ij}$, $M^{ij,+}$ and $N^{ij}$. We will
call the algebra (\ref{AdSGenA}) -- (\ref{AdSGenA3}) the symmetry
 algebra of interacting HS theory in AdS space-time.

The commutation relations above differ from the corresponding flat space ones (\ref{VGenA}),
in that they involve extra terms  which when acting on states give $O(1/L^2)$ contributions.
These terms are sub-leading in the $L\rightarrow \infty$ limit, hence the algebra (\ref{AdSGenA})
contracts to the flat space--time algebra (\ref{VGenA}) in the small curvature limit. This implies that free
HS gauge fields in flat space-time can be viewed as the zero curvature limit of free HS gauge fields on
AdS. However, the interacting HS gauge fields on AdS do not have a smooth $L\rightarrow\infty$ limit since
the interaction vertices contain positive powers of $L$. Nevertheless, as we shall see below, the functional
form of the cubic vertex of HS gauge fields on AdS differs from the cubic vertex in flat space-time by terms
which are sub-leading as $L\rightarrow\infty$.

From the explicit form (\ref{COMU1}) the
first term $[\nabla_\mu, \nabla_\nu]$ leaves the "scalar" Fock
state invariant and only the oscillator piece contributes, which
can be written in terms of the standard (\ref{VGen}) generators.
The algebra (\ref{AdSGenA})-(\ref{AdSGenA3}) acting on 
states becomes:
\begin{eqnarray}\label{AdSGenAA}
[l^{ij}, l^{mn,+}] &=& \delta^{im}l_{0}^{jn} + \frac{1}{L^2}
\delta^{jn}[ N^{mj}N^{mi} + (({\cal D}-1)\delta^{ij}-1)N^{mi}\\
&&-\frac{{\cal D}}{2} (\delta^{mi}+\delta^{mj})N^{ij} -
\frac{{\cal D}^2}{4}\delta^{mj}\delta^{mi}-4M^{mj,
+}M^{ij}]\nonumber
\end{eqnarray}
\begin{equation}\label{AdSGenAA2}
[l^{mn}, l^{kl}] = \frac{1}{L^2} \delta^{nl}[(\frac{{\cal
D}}{2}-1) (\delta^{kl} M^{ml}-\delta^{ml}M^{kl})
+N^{lm}M^{kl}-N^{kl}M^{ml}]
\end{equation}
\begin{eqnarray} \label{AdSGenAA4}
[l^{ij}_0, l^{mn}] &=& \frac{1}{L^2}\delta^{jn} (2l^{ji,+} M^{mn}
-N^{jm} l^{ni})+ \frac{1}{L^2}\delta^{in} (2l^{ij,+} M^{mn}
-N^{im} l^{nj}) \\ \nonumber &&+
\frac{1}{L^2}(\delta^{jn}\delta^{jm}l^{ni})(1- \frac{{\cal D}}{2})
+ \frac{1}{L^2}(\delta^{in}\delta^{im}l^{ni})(1- \frac{{\cal
D}}{2}) \\ \nonumber
&&-\frac{1}{L^2}(\delta^{in}\delta^{ij}l^{ni})(1-{\cal D})
\end{eqnarray}
\begin{eqnarray}\label{AdSGenAA3}
[l_0^{ij},l_0^{kl}]&=&
\frac{1}{L^2}\delta^{jk}(l^{ji,+}l^{jl}-l^{jl,+}l^{ji})+
\frac{1}{L^2}\delta^{ik}(l^{ij,+}l^{il}-l^{il,+}l^{ij})+ \\
\nonumber
&&\frac{1}{L^2}\delta^{jl}(l^{ji,+}l^{jk}-l^{jk,+}l^{ji})+
\frac{1}{L^2}\delta^{il}(l^{ij,+}l^{ik}-l^{ik, +}l^{ij})- \\
&&\frac{1}{L^2}({\cal
D}-1)(\delta^{jk}\delta^{jl}+\frac{1}{L^2}\delta^{ik}\delta^{il})l_0^{ij}+
\frac{1}{L^2}({\cal
D}-1)(\delta^{ik}\delta^{jk}l_0^{il}+\delta^{il}\delta^{jl}l_0^{ik})
\nonumber
\end{eqnarray}

We find it again useful to demonstrate how the calculations are done
 in the interacting case on AdS with an example.
\begin{eqnarray}\label{example2}
&&[l^{12}, l^{22,+}]\ l^{12,+}\varphi^1_\rho \alpha^{\rho,1
+}|0\rangle_1 \otimes \varphi^2|0\rangle_2=\alpha^{2,+}_\mu
D^2_{\mu \nu} \alpha^1_\nu  \ l^{12,+} \varphi^1_\rho \alpha^{\rho
1, +}|0\rangle_1 \otimes \varphi^2|0\rangle_2=
\nonumber \\
&&-\frac{1}{L^2}(l^{22,+} (N^{11}-1+\frac{{\cal D}}{2})-2M^{12,+}
l^{12}) \ \varphi^1_\rho \alpha^{\rho,1 +}|0\rangle_1 \otimes
\varphi^2|0\rangle_2= \nonumber \\
&&\frac{i}{L^2} \alpha^{2,+}_{\mu} \alpha^{1,+}_\nu({\cal D}
\varphi^\nu_1(\nabla^\mu \varphi_2) - g^{\mu \nu}\varphi_1^\rho
(\nabla_\rho \varphi_2))|0\rangle_1 \otimes |0\rangle_2
\end{eqnarray}
There is only a $p^2_\lambda$ from the second Hilbert space
involved in the example above. In the first equality we used
(\ref{AdSGenA}). In the second equality we acted with $D^2_{\mu
\nu}$ on the $p^2_\lambda$ of the $l^{12,+}$ operator using
(\ref{COMU1}) and (\ref{Cnabla}). This was the only "tensor"
operator in the 2nd Hilbert space, since $\varphi^2$ is a scalar.
Consequently, we commuted operators $\alpha^i_\sigma$ and
$p^2_\sigma$ past each other to bring the result to the second
line of (\ref{example2}). Finally we used the algebra of
(\ref{AdSGenAA})--(\ref{AdSGenAA3}) to complete the calculations
since no other "vector" operator, in the the 2nd Hilbert space,
was left for $l^{22+}$ or $l^{12}$ to act upon.

From the manipulations above we conclude the following: The
algebra of constraints being obviously more complicated than in
the case of flat space-time shares its main property-- namely it
preserves the polynomial form of (\ref{VExp}), (\ref{WExp}),
(\ref{XEXP}). Therefore we can proceed in an analogous manner as
in the flat case.

\subsection{BRST Invariance Constrains for the Cubic Vertex on AdS}

The next step is to choose an expansion of the cubic vertex in
terms of the AdS generators (\ref{hatpdef}) and (\ref{VGen}).
 In the AdS case the
creation generators of (\ref{LIVGen}) do not commute among each other, unlike the flat
case, as one can see from i.e. (\ref{AdSGenA4}). Nevertheless,
 we can choose a ${\it standard}$ ordering as in (\ref{XEXP}). All other possible
orderings can be brought in the
${\it standard}$ form ( i.e.,
use an analogue of the Weyl ordering in quantum mechanics),
using the algebra
(\ref{AdSGenA}-\ref{AdSGenA3}) and the manipulations described in the previous
subsection, modulo $\frac{1}{L^2}$ terms \footnote{The action of
$D^i_{\mu \nu}$ on "tensors" produces terms proportional to $\frac{1}{L^2}$ as on can
easily verify from (\ref{COMU1})
and (\ref{Cnabla}).}
that affect lower dimension terms in the $L^2$ expansion of $X^n$.
These latter terms can again
be brought in the ${\it standard}$ form following the same procedure and finally be
absorbed in the definition of the matrix elements with lower dimension than the one we
started from.

In addition although naively we do not have momentum conservation
in AdS space-time, we can still make use of the equation $\sum_i
p^\mu_i=0$, since it leads into total derivative terms in
the Lagrangian.

To conclude, one can construct the same linearly independent set
of generators as in (\ref{LIVGen}). The expansion of the
coefficients is exactly the same as in (\ref{XEXP}) with all
generators the AdS equivalent of the flat ones. Using the explicit
form of (\ref{QAdS}) it is straightforward to write down the
equations resulting from (\ref{VBRST}). They are the same as in
flat case with the substitution $l_0 \to \hat{l}_0$ as in
(\ref{l0AdS}) and some modifications analogous to those in
(\ref{dVAdS}). The final result is:
\begin{eqnarray}\label{VAdSEQN}
&&c^{i,+}[l^i X^1-l^{s,+} X^2_{is}-\hat{l}_0^sX^3_{is}
+\frac{16}{L^2}M^{s,+}X^5_{is;ss}]=0  \nonumber \\
&&c^{i,+} \gamma^{ jk,+}[l^iX^2_{jk}-2l^{s,+}X^4_{(is);(jk)}
-\hat{l}_0^s X^5_{jk;is}
 \nonumber \\
&&+\frac{8}{L^2}(\delta_{jk} M_j X^3_{jk} -6 M^{s,
+}X^8_{(ss);(jk);is})]=0
 \\
&&c^{i,+} \beta^{jk,+}[-\delta_{jk}X^2_{ij}+ l^i
X^3_{jk}-l^{s,+}X^5_{is;jk}-2\hat{l}_0^sX^6_{(is);(jk)}
-\frac{32}{L^2}M^{s, +}X^9_{ss;(jk);(is)}]=0 \nonumber
\end{eqnarray}
\begin{eqnarray}
&&c^{i,+} \gamma^{jk,+}\gamma^{lm,+}[l^iX^4_{(jk);(lm)}- 3l^{s,+}X^7_{(is);(jk);(lm)}
-\hat{l}_0^sX^8_{(jk);(lm);is} \nonumber \\
&&-\frac{8}{L^2}\delta_{jk} M^j
X^5_{lm;ij}]=0 \nonumber \\
&&c^{i,+} \gamma^ {jk,+}\beta^{ lm,+}[-2\delta_{lm}X^4_{(il);(jk)}+
l^iX^5_{jk;lm}-2l^{s,+}X^8_{(is);(jk);lm}-2\hat{l}_0^sX^9_{jk;(is);(lm)} \nonumber \\
&&+\frac{16}{L^2} \delta_{jk} M^j X^6_{(lm);(ij)}]=0
\nonumber \\
&&c^{i,+} \beta^{jk,+}\beta^{lm,+}[-\delta_{jk} X^5_{ji;lm}+ l^i
X^6_{(jk);(lm)}-l^{s,+}X^9_{is;(jk);(lm)}-3\hat{l}_0^sX^{10}_{(is);(jk);(lm)}]=0 \nonumber
\end{eqnarray}
 Combinations involving the operator $\hat{l}_0^s$
should be understood as follows. For example the term in the first equation
$c^{i+}\hat{l}_0^sX^3_{is}$ is  a result of an  action of the operator
$c^{i +}_0 \hat{l}_0^i$ at $X^3_{mn} \beta^{+mn}$ and using the expression
(\ref{l0AdS})
\begin{eqnarray}\label{l0AdS11}
c^{i,+}_0 \hat{l}_0^i X^3_{mn} \beta^{mn, +}&=&-c^{i,+}
 ({p}^{\mu, s} {p}^s_\mu + \frac{1}{L^2}({(\alpha^{\mu, s +} \alpha^s_\mu)}^2 + {\cal
D}\alpha^{\mu, s +} \alpha^s_\mu -6 \alpha^{\mu, s +} \alpha^s_\mu
-  \\ \nonumber &&{(2{\cal D} -6)}^s -4 M^{s,+} M^s) X^3_{is} -
 \frac{1}{L^2}c^{i,+}(4 \alpha^{\mu, i +} \alpha^i_\mu + {(2{\cal D} -6)}^i) X^3_{ii}.
 \nonumber
\end{eqnarray}
The equations in (\ref{VAdSEQN}) are  more difficult to
analyze compared to flat case despite their apparent similarity.
  The main reason is obvious from the algebra
(\ref{AdSGenA})--(\ref{AdSGenA3}) which has nontrivial commutators
containing $D^i_{\mu \nu}$. This causes more a technical
difficulty rather than a conceptual one. It would be intersting to
find a solution in a closed compact form (if such a solution
exists of course) but at the present moment we are content to have
a well defined iteration procedure and a system of equations which
can be straightforwardly solved via this procedure.

\setcounter{equation}0\section{Summary and Outlook}

In the present paper we have addressed the problem of constructing
the cubic interaction vertex  of Higher Spin Theory in the
``metric-like formalism`` on ${\cal D}$ -- dimensional flat and
AdS spaces. We have discussed the free equations of motion for
Higher Spin fields in flat and AdS space-times in triplet
formalism. The only principle we have followed in this
construction is the requirement of gauge invariance of the
Lagrangian. These equations describe reducible representation of
the Poincare and AdS groups. To obtain an irreducible
representation on has to add certain off shell constraints to the
field equations.

Assuming the cubic interaction vertex to be a series in ghost and
oscillator variables we have obtained the equations which
determine the vertex. We outlined the way how these equations can
be solved level by level in oscillator expansions.
  The vertex obtained in this way contains  a part which produces `fake interactions``
i.e. the ones which can be obtained form the free field Lagrangian
via field redefinitions. We have shown how in practice this
trivial part of the vertex can be factorized out by solving the
cohomologies of the corresponding BRST operator which determines a
free part of the Lagrangian. As a result, the gauge invariant
formulation  the Lagrangian contains alongside  physical modes
some auxiliary fields as well. Finding the form of the vertex is
essentially based on the symmetry algebras of interacting HS
fields  in flat (\ref{VGenA}) and AdS
(\ref{AdSGenA})-(\ref{AdSGenA3}) space-times.

There are several open problems to address namely
\begin{itemize}

\item Our results can used to focus on particular sets of fields,
having in mind holography. In particular, it would be quite
interesting to reproduce holographically the known results for the
conformal 3-point functions of fields with spin 1 and 2 \cite{PO}.

\item A
crucial test would be  is to compare the cubic interaction vertices obtained in the
present approach to the ones obtained in a ``frame --like ``
formulation by Vasiliev   \cite{Vasiliev:1990en}.

\item Our approach can be used to discuss the interactions of HS
fermionic fields, as well as of HS fields represented by tensors
with mixed symmetry.

\item  Finally, it would be very desirable if our calculations shed some light into
the conjectured link of HS gauge theory with the high energy  behavior  of
string theory in flat \cite{Moeller:2005ez} -- \cite{Gross:1987ar}
and in AdS spacetimes.

\end{itemize}

\vspace{1cm}

\noindent {\bf Acknowledgments.} It is a pleasure to thank X.
Bekaert, C. Coriano, N. Irges, A. Koshelev, P.~ Lavrov, M. Matone,
P. Pasti, D. Sorokin, P. Sundell, M. Tonin, T. Tomaras, M.
Vasiliev, P. West and R. Woodard for useful discussions. This work
was partially supported by the INTAS grant, project
INTAS-03-51-6346. The work of I.L.B was partially supported by
RFBR grant, project No 06-02-16346, joint RFBR/DFG grant, project
No 06-02-04012, DFG grant, project No 436 RUS 113/669/0-3 and
grant LRSS, project No 4489.2006.2. The work of A. C. P. was
partially supported by the PYTHAGORAS II Research Program
K.A.2101, of the Greek Ministry of Higher Education. The work of
A.F. is supported by a PYTHAGORAS Research Program K.A.1955 of the
Greek Ministry of Higher Education and partially supported by the
European Commission, under RTN program MRTN-CT-2004-0051004 and by
the Italian MIUR under the contract PRIN 2005023102. The work of
M.T. was supported by the European contract MRTN-CT-2004-512194.


\renewcommand{\thesection}{A}

\setcounter{equation}{0}

\renewcommand{\theequation}{A.\arabic{equation}}

\section{Field Redefinitions in BRST Formalism}\label{ApB}

In this appendix we will demonstrate how one can use the FR
freedom $|W\rangle$ to eliminate `` fake interactions`` and to
bring $|V \rangle$ in a convenient form.

\subsection{Equations for Field Redefinitions on Flat Space - Time}
A direct computation of (\ref{VFR}) using (\ref{QBRST}), (\ref{VExp}) and
(\ref{WExp}) leads to the following transformations
 for the expansion
coefficients of the vertex:
\begin{eqnarray}\label{dVExp}
&&\delta X^1= l^{i,+}W^1_i+ l_0^i W^2_i  \\
&&[\delta X^2_{ij}] \gamma^{ij,+}= [l_i W^1_j + 2l^{s,+}
W^3_{s;ij}-l_0^sW^4_{i;js}]\gamma^{ij,+}
\nonumber \\
&&[\delta X^3_{ij}] \beta^{ij,+}=[\delta_{ij} W^1_i
+l_iW^2_j+l^{s,+}W^4_{s;ij}+2l_0^sW^5_{s;ij}]\beta^{ij,+} \nonumber \\
&&[\delta X^4_{(ij);(kl)}] \gamma^{ij,+} \gamma^{kl,+}=[l_{i}W^3_{j;(kl)} +
3l^{s,+}W^6_{s;(ij);(kl)}+l^{s}_0W^7_{l;(ij);ks}] \gamma^{ij,+} \gamma^{kl,+} \nonumber \\
&&[\delta X^5_{ij;kl}] \gamma^{ij,+} \beta^{kl,+}=[2\delta_{kl}W^3_{l;ij}+ l_iW^4_{j;kl}+
2l^{s,+}W^7_{s;ij;kl}-2l_0^sW^8_{j;(kl);(is)}]\gamma^{ij,+} \beta^{kl,+}  \nonumber \\
&&[\delta X^6_{(ij);(kl)}]\beta^{ij,+} \beta^{kl,+} =[\delta_{(ij)}W^4_{j;(kl)}
+l_{(i}W^5_{j);kl}
+l^{s,+} W^8_{s;(ij);(kl)} \nonumber \\
&&+3l_0^s
W^9_{s;(ij);(kl)}] \gamma^{ij,+} \beta^{kl,+} \nonumber \\
&&[\delta X^7_{(ij);(kl);(mn)}] \gamma^{ij,+} \gamma^{kl,+} \gamma^{mn,+}  = [l_{i}W^6_{j;(kl);(mn)}-l_0^s
W^{10}_{n;(ij);(kl);ms}]\gamma^{ij,+} \gamma^{kl,+} \gamma^{mn,+}  \nonumber \\
&&[\delta X^{8}_{(ij);(kl);mn}] \gamma^{ij,+} \gamma^{kl,+} \beta^{mn,+} =[3\delta_{mn}W^5_{n;(ij);(kl)}+
l_{i}W^7_{j;(kl);mn} \nonumber \\
&&+3l^{s, +}W^{10}_{s;(ij);(kl);mn}-2l_0^sW^{11}_{l;(ij);ks;mn}]
\gamma^{ij,+} \gamma^{kl,+} \beta^{mn,+}
\nonumber \\
&&[\delta
X^9_{ij;(kl);(mn)}]\gamma^{ij,+} \beta^{kl,+} \beta^{mn,+}
=[2\delta_{kl}W^7_{l;ij;(mn)}+l_iW^8_{j;(kl);(mn)} \nonumber \\
&&+ 2l^{s,+}W^{11}_{s;ij;(kl);(mn)}
-3l_0^sW^{12}_{j;is;(kl);(mn)}]\gamma^{ij,+} \beta^{kl,+}
\beta^{mn,+}
\nonumber \\
&&[\delta X^{10}_{(ij);(kl);(mn)}]\beta^{ij,+} \beta^{kl,+} \beta^{mn,+}=[-\delta_{ij}W^8_{j;(kl);(mn)}+
l_{i}W^9_{j;(kl);(mn)} \nonumber \\
&&+ l^{s,+}W^{12}_{s;(ij);(kl);(mn)}]\beta^{ij,+} \beta^{kl,+}
\beta^{mn,+} \nonumber
\end{eqnarray}

In order to analyze these equations we need to determine the
action of operators $l^i,l^{i,+},l_0^i$ on the matrix elements
$W^l$. First let us write down the commutator relations for the
basis in (\ref{LIVGen}):
\begin{eqnarray}\label{CLIVGen}
&[l^i,l^{j,+}]&= \delta^{ij}l_0^j \nonumber \\
&[l^i,I^{j,+}]&=-\frac{1}{2}\delta^{ij}
\epsilon_{ikl}(l_0^k-l_0^l)
\\
&[l_i, M^{kl,+}]&= \frac{3}{2} \delta^{ik}\delta^{kl} l^{k,+}
-\frac{1}{4}(\delta^{li}l^{k,+}+\delta^{ki}l^{l,+})  +
\frac{1}{4}(\delta^{li}\epsilon_{klm}I^{k,+}+\delta^{ik}\epsilon_{lkm}I^{l,+})
\nonumber
\end{eqnarray}
with all other commutators vanishing. The indexes $i,j,k$ run over
the three Hilbert spaces. Based on the algebra above we can deduce
the following set of simple transformation rules:
\begin{eqnarray}\label{Trans}
&&l^{0} W(n_i:n_i) \rightarrow  W(n_i:n_i+1) \nonumber \\
&&l^{+i} W(m_i:m_i)  \rightarrow    W(m_i:m_i+1) \\
&&\nonumber \\
&&l^i W(n_i:n_i,m_i:m_i,k_i:k_i,p_i:p_i,r_{ij}:r_{ij})
\longrightarrow
\nonumber \\
&& m_i W(n_i:n_i+1,m_i:m_i-1) + p_i W(m_i:m_i+1,p_i:p_i-1)
\nonumber \\
&&-\frac{k_i}{2}  \epsilon_{ikl}  (
W(n_k:n_k+1,k_i:k_i-1)-W(n_l:n_l+1,k_i:k_i-1))
\nonumber \\
&&-\frac{1}{4} \sum_l r_{il} W(r_{il}:r_{il}-1,m_l:m_l+1)
+\frac{1}{4} \sum_l r_{il} \epsilon_{lim}
W(r_{il}:r_{il}-1,k_l:k_l+1) \nonumber
\end{eqnarray}
Let us explain our notation in the above equation. We use $W(n_i:n_i)$
for an element which has an expansion as in (\ref{XEXP}): $W_{n_i,
\dots} (l_0^i)^{n_i} \dots$.
Then the transformation law in the
second equation above means that we have on the RHS the expansion:
$W(n_i:n_i+1)=W_{n_i, \dots} (l_0^i)^{n_i+1} \dots$.
Analogously are defined the other quantities in (\ref{Trans}).

As we have  already mentioned  the $l_0^i$ generators act on states of
the ith-Hilbert space as
flat space-time Laplacian. In addition the $l^{+i}$ operators act
on the left, on bra states, as divergences resulting in
$\nabla^\mu \Phi_{\mu \dots}$ terms.
We will show that an appropriate choice of the FR
functions $W^n$, can be used to eliminate most of the dependence,
of the $X^n$ coefficients, on these operators.

~From ~the ~first ~of ~(\ref{dVExp}) ~and ~(\ref{Trans}) ~we ~can ~easily
~see ~that ~using ~all ~of ~the ~$W^1_{1;n_i \geq 0, m_i\geq 0}$ ~freedom
~we ~can ~eliminate ~the ~$m_1>0$ terms in $X^1$ and we are left with
the element $X^1_{n_i \geq 0, m_1=0, m_{2,3} \geq 0}$. We can go on and use
$W^1_{2;n_i \geq 0, m_1=0, m_{2,3} \geq 0}$ to eliminate $m_2>0$
terms. Next we use some of the $W^1_3$ freedom to restrict to
$X^1_{n_i \geq 0, m_i=0}$. Notice that in the second and third
steps we have not used all the $W^1_{2,3}$ freedom available.
Proceeding in a similar fashion we can show that some of the
$W^2_i$ freedom can be used to eliminate all $n_i>0$ matrix
elements of $X^1$. Working carefully with the remaining of
(\ref{dVExp}) we can eliminate most of the $n_i>0$ and $m_i > 0$
dependence of the $X^n_{\dots}$ . The non-vanishing matrix
elements for the specific scheme chosen are:
\begin{eqnarray}\label{Vcomp}
&&X^1_{n_i=0,m_i=0 \dots} \ \ X^2_{n_i=0,m_i=0\dots} \ \
X^3_{n_i=0,m_i\geq0}
\nonumber \\
&&X^4_{n_i=0,m_i=0\dots} \ \ X^5_{n_i=0, m_i \geq 0,} \ \
X^6_{n_i=0,m_i \geq 0}
\nonumber \\
&&X^7_{n_i=0,m_i\geq 0\dots} \ \ X^8_{n_i=0,m_i \geq 0\dots}
\ \ X^9_{n_i=0,m_i
\geq 0\dots}\nonumber \\
&&X^{10}_{n_i\geq 0, m_i \geq 0\dots}
\end{eqnarray}

Another useful FR scheme is the following
\begin{eqnarray}\label{VcompB}
&&X^1_{n_i=0,m_i=0 \dots} \ \ X^2_{n_i \geq 0,m_i=0\dots} \ \
X^3_{n_i=0,m_i=0}
\nonumber \\
&&X^4_{n_i \geq 0,m_i=0\dots} \ \ X^5_{n_i \geq 0, m_i = 0,} \ \
X^6_{n_i=0,m_i = 0}
\nonumber \\
&&X^7_{n_i \geq 0,m_i\geq 0\dots} \ \ X^8_{n_i \geq 0,m_i =0\dots}
\ \ X^9_{n_i \geq 0,m_i= 0\dots}\nonumber \\
&&X^{10}_{n_i \geq 0, m_i =0 \dots}
\end{eqnarray}

We should emphasize that there are various FR schemes like  (\ref{Vcomp}) and
(\ref{VcompB}).
 We have chosen
the specific scheme because we believe it is a very economical
off-shell Lagrangian action where most redundant  terms
are absent. Obviously two interaction vertices which differ by a
FR are equivalent on-shell.

\subsection{Equations for Field Redefinitions on AdS}
For AdS FR equations can be computed in an analogous way as for
flat case. Equation (\ref{brst}) can be written in the following
more compact form:
\begin{equation}\label{QAdS}
Q =
c_0\hat{l}_0 +c l^{+}+c^+ l -\frac{8}{L^2} c_0(\gamma^+ M + \gamma M^+) - c^+cb_0
\end{equation}
With the BRST charge written in this form it is straightforward to
compute the analogous of (\ref{dVExp}) for the AdS case. We only
need to substitute
\begin{eqnarray}\label{l0AdS}
l_0 \rightarrow \hat{l}_0&=& {p}^\mu {p}_\mu +
\frac{1}{L^2}({(\alpha^{\mu +} \alpha_\mu)}^2 + {\cal
D}\alpha^{\mu +} \alpha_\mu -6 \alpha^{\mu +} \alpha_\mu -2{\cal
D} +6
-4 M^+ M +\nonumber \\
&& c^+b (4 \alpha^{\mu +} \alpha_\mu +2{\cal D} -6 ) + b^+c (4 \alpha^{\mu +} \alpha_\mu +2{\cal D} -6) +
12 c^+b b^+c)
\end{eqnarray}
in (\ref{dVExp}) and compute the only modification coming from the
third term of the last line in (\ref{QAdS}). The final result is
\begin{eqnarray}\label{dVAdS}
&&\delta X^{1;AdS}= \delta X^1  -\frac{8}{L^2} M^{s,+} W^4_{s;ss}\nonumber \\
&&[\delta X^{2;AdS}_{ij}] \gamma^{ij,+}=[\delta
X^2_{ij}-\frac{8}{L^2}(\delta_{ij}M_iW^2_i + 4 M^{s,+}
W^7_{s;ij;ss})] \gamma^{ij,+}
\nonumber \\
&&[\delta X^{3;AdS}_{ij}]\beta^{ij,+}= [\delta X^3_{ij}-\frac{32}{L^2}M^{s,+}W^8_{s;(ss);(ij)}]\beta^{ij,+}\nonumber \\
&&[\delta X^{4;AdS}_{(ij);(kl)}]\gamma^{ij,+}\gamma^{kl,+}= [\delta X^4_{(ij);(kl)}+
\frac{8}{L^2}  (\delta_{ij} M_iW^4_{l;ki}
-9M^{s,+}W^{10}_{s;(ij);(kl);ss})]\gamma^{ij,+}\gamma^{kl,+}\nonumber \\
&&[\delta X^{5;AdS}_{ij;kl}]\gamma^{ij,+}\beta^{kl,+} =[\delta
X^5_{ij;kl}-\frac{16}{L^2}(\delta_{ij}M_iW^5_{i;kl} + 6
M^{s,+}W^{11}_{s;ij;(kl);(ss)})] \gamma^{ij,+}\beta^{kl,+}
\nonumber \\
&&[\delta X^{6;AdS}_{(ij);(kl)}] \beta^{ij,+}\beta^{kl,+}=[\delta X^6_{(ij);(kl)} -\frac{72}{L^2}
M^{s,+}W^{12}_{s;(ss);(ij);(kl)}]\beta^{ij,+}\beta^{kl,+} \\
&&[\delta X^{7;AdS}_{(ij);(kl);(mn)}]\gamma^{ij,+}\gamma^{kl,+} \gamma^{mn,+}=[\delta
X^7_{(ij);(kl);(mn)} \nonumber \\
&&+\frac{8}{L^2} \delta_{ij} M_i
W^7_{l;mn;ki}]\gamma^{ij,+}\gamma^{kl,+} \gamma^{mn,+}
  \nonumber \\
&&[\delta X^{8;AdS}_{(ij);(kl);mn}]\gamma^{ij,+}\gamma^{kl,+} \beta^{mn,+}=[\delta X^{8}_{(ij);(kl);mn}
\nonumber \\
&&+\frac{16}{L^2} \delta_{ij} M_iW^8_{l;(ki);(mn)}]\gamma^{ij,+}\gamma^{kl,+} \beta^{mn,+}
\nonumber \\
&&[\delta X^{9;AdS}_{ij;(kl);(mn)}]\gamma^{ij,+}\beta^{kl,+} \beta^{mn,+}=[\delta
X^9_{ij;(kl);(mn)} \nonumber \\
&&-\frac{24}{L^2} \delta_{ij}M_i
W^9_{i;(kl);(mn)}]\gamma^{ij,+}\beta^{kl,+} \beta^{mn,+}
\nonumber \\
&&[\delta X^{10;AdS}_{(ij);(kl);(mn)}]\beta^{ij,+}\beta^{kl,+} \beta^{mn,+} = [\delta
X^{10}_{(ij);(kl);(mn)}]\beta^{ij,+}\beta^{kl,+} \beta^{mn,+}\nonumber
\end{eqnarray}

\end{document}